\documentclass[10pt]{article}
\usepackage{latexsym}
\usepackage{amsmath}
\usepackage{pstricks} 

\long\def\ca#1\cb{} 

\newcommand{\becs}{\begin{cases}}
\newcommand{\bem}{\begin{matrix}}

\newcommand{\encs}{\end{cases}}
\newcommand{\enm}{\end{matrix}}

\newcommand{\lra}{\leftrightarrow }

\newcommand{\OR}{\mbox{\small OR}}

\newcommand{\ra}{\rightarrow }

\def\outl#1{\par{\medskip\noindent\hspace*{.5cm}\bf
      \mathversion{bold}#1\mathversion{normal}\smallskip} }
 \def\xa{} \def\xb{}  

 \def\outl#1{}  \def\xa{} \def\xb{}  

\ca
 \def\outl#1{\par{\medskip\noindent\hspace*{.5cm}\bf
      \mathversion{bold}#1\mathversion{normal}\smallskip} }
 \long\def\xa#1\xb{}
 
\cb

\begin{document}

\title{Reply to Comment [axXiv:1610.07734] by L. Vaidman on\\ 
``Particle path through a nested  Mach-Zehnder interferometer''{\\}
 [R.\ B.\ Griffiths, Phys.\ Rev.\ A 94 (2016) 032115]}
\author{Robert B. Griffiths
\thanks{Electronic address: rgrif@cmu.edu}\\ 
Department of Physics\\
Carnegie Mellon University\\ Pittsburgh, PA 15213} 
\date{Version of 10 April 2017}

\maketitle

\begin{abstract} 	
  The correctness of the consistent histories analysis of weakly interacting
  probes, related to the path of a particle, is maintained against the
  criticisms in the Comment, and against the alternative approach described
  there, which receives no support from standard (textbook) quantum mechanics.
\end{abstract}


\xb \outl{Nested Mach-Zehnder with beam splitters and detectors} \xa

The Comment \cite{Vdmn17} deals with an analysis \cite{Grff16} of the path
followed by a photon (hereafter a ``particle'') while inside the nested
Mach-Zehnder interferometer shown in the figure. The photon enters through
channel $S$ on the left, encounters various beam splitters leading to channels
labeled $A$, $D$, etc., and is then detected by one of three detectors. A key
feature is that the beamsplitters in the inner Mach-Zehnder containing channels
$B$ and $C$ are such that a photon which enters through $D$ will always emerge
in $F$, never in $E$.

\begin{figure} [h]
$$
\begin{pspicture}(-1,-1.5)(6.5,3.5) 
\newpsobject{showgrid}{psgrid}{subgriddiv=1,griddots=10,gridlabels=6pt}
\def\lwd{0.035} 
\def\lwb{0.10}  
\def\lwn{0.015}  
\psset{unit=0.7cm, 
labelsep=2.0,
arrowsize=0.150 1,linewidth=\lwd}
\def\rdet{0.35}  
\def\detect{
\psarc[fillcolor=white,fillstyle=solid](0,0){\rdet}{-90}{90}
\psline(0,-\rdet)(0,\rdet)}
\def\drad{0.35}
\def\detectd{
\psarc[fillcolor=white,fillstyle=solid](0,0){\drad}{180}{0}
\psline(-\drad,0)(\drad,0) }
\def\bsw{0.3}
\def\bsp{\psline[linewidth=\lwd](-\bsw,\bsw)(\bsw,-\bsw)}
\def\mrw{0.35}
\def\mrw{0.3}\def\mrwm{0.2}\def\mrwp{0.4}
\def\mirrorv{
\psline[linewidth=\lwd]%
(\mrw,-\mrw)(\mrwm,-\mrwp)(-\mrwp,\mrwm)(-\mrw,\mrw)(\mrw,-\mrw)}
\def\mirrore{
\psline[linewidth=\lwd]%
(\mrw,-\mrw)(\mrwp,-\mrwm)(-\mrwm,\mrwp)(-\mrw,\mrw)(\mrw,-\mrw)}
\psline(-1,3)(4,3)(4,-1.5)
\psline(0,3)(0,0)(5.5,0)
\psline(2,3)(2,1.5)(5.5,1.5)
\rput(0,3){\bsp} \rput(2,3){\bsp} \rput(4,1.5){\bsp} \rput(4,0){\bsp}
\rput(0,0){\mirrorv} \rput(2,1.5){\mirrorv} \rput(4,3){\mirrore}
\rput(5.5,1.5){\detect} \rput(5.5,0){\detect} \rput(4,-1.1){\detectd}
\rput[b](-0.8,3.2){$S$} \rput[b](1,3.2){$D$} \rput[l](4.1,0.7){$E$}
\rput[b](1,0.2){$A$} \rput[b](2.5,1.7){$B$} \rput[t](3.5,2.8){$C$}
\rput[b](5.0,1.7){$H$} \rput[b](5.0,0.2){$G$} \rput[l](4.1,-0.7){$F$}
\rput[l](4.4,-1.3){$D_1$} \rput[l](5.9,0){$D_2$} \rput[l](5.9,1.5){$D_3$}
\end{pspicture}
$$
\caption{%
  Nested Mach-Zehnder interferometer. The tilted solid lines are beam splitters;
  the double tilted lines are mirrors; the semicircles are
  detectors. The horizontal and vertical lines indicate different channels
  which are possible particle (photon) paths. }
\label{fgr1}
\end{figure}
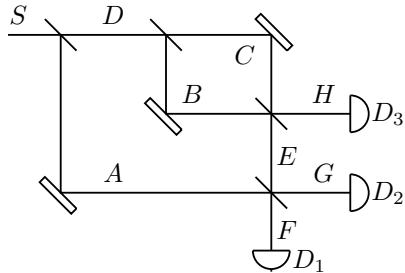

\xb \outl{``Weak trace'' vs CH re prior location of particle detected in $D_1$}
\xa

Of particular interest is the case in which the particle is detected in $D_1$.
In \cite{Vdmn13} it was claimed that this particle was earlier in all three
channels $A$, $B$, and $C$ but, somewhat mysteriously, not in $D$ or $E$. That
differs from the conclusion in \cite{Grff16}, in agreement with \cite{LAAZ13},
that such a photon was earlier in $A$ and not elsewhere. These contrary claims
reflect different ways of analyzing the situation. In \cite{Grff16} the
analysis was based on the consistent histories (CH) interpretation of quantum
mechanics, for which the standard reference is \cite{Grff02c}, while
\cite{Grff14b} contains a short introduction. By contrast, \cite{Vdmn13}
employs a phenomenological principle which says that a particle can be said to
have been in any location where it left a ``weak trace.'' One way to model a
weak trace is by means of a probe which is weakly perturbed by the presence of
a particle in a channel. This approach was employed in Sec.~V.B of
\cite{Grff16} using two-state or qubit probes $a$, $b$, etc.\ coupled to
channels $A$, $B$, etc., along with a special probe $w$ coupled to the
combination $B+C$; and it was claimed that the statistics of measurement
outcomes on these probes supported the analysis based on a situation with no
probes present, that the particle detected in $D_1$ was earlier in $A$ and not
elsewhere. It is this claim that is contested in \cite{Vdmn17}.

\xb \outl{CH essentials: Not measurement based; Qm properties $\lra$ subspaces;
SFR} \xa

It should be noted that in the CH approach, unlike standard quantum mechanics
(SQM) understood as what one finds in typical textbooks, \emph{measurement}
plays no fundamental role. Measurements are simply physical processes which can
be interpreted by means of quantum concepts that apply to all microscopic or
macroscopic quantum processes. Thus the infamous measurement problem of quantum
foundations is completely missing from (or fully resolved by) CH, which also
has no need of wavefunction collapse, though that can sometimes be used as a
mathematical tool to obtain results accessible by other less confusing methods
whose physical content is much clearer. In addition, CH provides the tools
needed for using measurement outcomes (``pointer positions'') to infer
something about the state of the microscopic measured system \emph{prior} to
when the measurement took place, thus resolving what can be called the ``second
measurement problem'' \cite{Grff15}. This is done by introducing a consistent
family of quantum histories, a \emph{framework}, in which prior microscopic
properties are represented by subspaces of the quantum Hilbert space or,
equivalently, projectors onto these subspaces. The numerous paradoxes which
beset SQM are avoided by use of the \emph{single framework rule}, which allows
the use of different frameworks, but prohibits combining results from different
incompatible frameworks. As applied to the situation shown in the figure, there
is framework in which a particle detected by $D_1$ was earlier in $A$ and not
in $B+C$, where $A$, $B$, and $C$ are projectors on particle states in the
corresponding channels. Note that the quantum projector $B+C$ does not mean
quite the same things as ``$B$ \OR\ $C$'', since a coherent superposition of
two wave packets, one in $B$ and one in $C$, will lie in the subspace $B+C$,
but not in $B$ or $C$ separately. No phenomenological principles are required
for this analysis, since the CH formalism itself provides all the tools needed
to give precise answers to questions about earlier particle locations.

\xb \outl{Incompatible CH frameworks: $D_1 \ra A$ or $\ra C$, but not both.}
\xa

For a special choice of parameters in \cite{Grff16} there is both a framework
which allows the inference from detection in $D_1$ to the particle having been
earlier in $A$, and a second framework, incompatible with the first, which can
be used to infer, again from detection in $D_1$, that the particle was earlier
in $C$. This is mentioned in \cite{Vdmn17}, but without noting the important
fact that these incompatible frameworks cannot be combined. In discussing a
particular run, one or the other framework provides a possible description, but
not both together, and this excludes reaching the conclusion that the particle
was both in $A$ and in $C$. (For a detailed analysis of this ``three-box
paradox'' see Sec.~22.5 of \cite{Grff02c}.) Thus when properly analyzed this
example provides no support for the claim in \cite{Vdmn13} that in a single
experimental run the particle can have been in several locations.

\xb \outl{CH interpretation of probes} \xa

A major difference between \cite{Vdmn17} and \cite{Grff16} is the
interpretation of the probes. In \cite{Grff16}, after the probes interact with
the particle they are measured later, after the run is over (and the particle
has been detected in one of the detectors), in an appropriate basis to
determine whether or not they have been triggered. If the measurement indicates
the probe has been triggered, this can be used to infer using CH (but not SQM)
that the particle was earlier present in the corresponding channel; e.g., it
was in $B$ if probe $b$ has been triggered. However, if the probe has not been
triggered this tells one very little: the particle may have been absent, or it
may have present but because the interaction was so weak it failed to trigger
the probe. (Think of the weak light source in Feynman's discussion in Ch.~1 of
\cite{FyLS65} of the two hole (double slit) experiment.) As noted previously,
CH makes no use of wavefunction collapse, and nothing in \cite{Vdmn17} should
be understood as indicating otherwise.

\xb \outl{Importance of coincidences among the probes} \xa

The claim in \cite{Vdmn17} that coincidences do not provide useful information
seems odd. If in a particular run the presence of a particle at a succession of
points is indicated by reliable measurements---they may be weak, but they do
not yield false positives---most experimental physicists would consider this as
evidence that the particle followed a particular path, especially with
detectors triggered in the proper temporal sequence, a situation obviously not
practical in the case of a photon, but easily arranged in a gedanken experiment
like the one in the figure by adding more sophisticated probes connected to
timers. Granted, in the absence of an appropriate theory of weak measurements
one can infer nothing from coincidences, but CH provides a consistent theory
whose results, reported in \cite{Grff16}, deserve to be taken seriously.

\xb \outl{Vd interpretation of probes compared to CH interpretation} \xa

By contrast, in \cite{Vdmn17} the probe system is analyzed using collapse of
the total unitarily evolved wavefunction for the particle and the probes at a
time when the particle, no longer in the Mach-Zehnder, is detected by $D_1$.
The result is a superposition state of the system of probes including cases in
which zero, one, two, etc. probes have triggered, with different amplitudes. No
interpretation of such a superposition is provided by SQM; instead, one assumes
that some sort of measurement will take place later, and the state just
mentioned can assign probabilities to the outcomes. These probabilities for the
outcomes of measurements on the probes are exactly the same as those in
Sec.~5.2 of \cite{Grff16} for the case of a particle emerging in channel $F$,
so CH and SQM agree on this. But in addition CH allows one to draw the
consequence that when, for example, probe $b$ triggered, the particle actually
was at the earlier time, when it interacted with the probe, in channel $B$. For
many physicists, especially those who carry out real experiments, this will
seem intuitively obvious or at least plausible, despite the fact that SQM as
found in the textbooks provides no justification for it. However, the statement
in \cite{Vdmn17} that measuring the probe can ``collapse the state of the
particle to the path of the probe,'' which is probably intended to mean that
the particle state is collapsed to the channel associated with the probe, is
not supported by CH, which makes no use of collapse, or by SQM, where collapses
do not somehow cause events in the past.

\xb
\outl{Triggered probes don't tell what happens when probes NOT triggered}
\xa

The concern expressed in \cite{Vdmn17}, that what happens in a run when one or
several probes are triggered need not represent the situation when no probes
are triggered, must be taken seriously. Looking at a small sample and
extrapolating the results to a much larger ensemble is common scientific
practice; think of measuring the half-life of some nuclear species and then
using the results for radioactive dating. Doing this requires making some
assumptions and a modicum of theory. In the case of quantum measurements there
is the difficulty that these can seriously disturb the measured system. Hence
it is worth noting that the theoretical structure provided by CH can make quite
definite statements about the position of a particle inside the interferometer,
conditioned on where it was finally detected, \emph{without} any reference to
weak measurements; such an analysis constitutes the bulk of \cite{Grff16}. Weak
measurements with probes are analyzed in only one section of that paper, where
they are shown to be perfectly consistent with the earlier results obtained
when no probes are present.

\xb 
\outl{Vd approach DEPENDS on weak measurements} \xa

\xb
\outl{Vd's incorrect conclusion from ignoring effect of $b$ when triggered}
\xa

By contrast, the weak trace approach in \cite{Vdmn13}, defended in
\cite{Vdmn17}, is heavily dependent on a certain interpretation of weak
measurements by probes which, as noted above, receives no support in SQM. So
the issue of deciding what happens when a probe is not triggered, or when there
are no probes to be triggered, is a serious problem. Indeed, the analysis in
Lao and \cite{Grff16} identifies a specific way in which the weak trace method
of \cite{Vdmn13} has led to an incorrect conclusion. When the $b$ probe
associated with $B$ is triggered it has a drastic effect: the particle which
would otherwise have emerged in $H$ has a significant probability of exiting
the inner Mach-Zehnder in channel $E$ and continuing on through $F$ to be
detected by $D_1$. This effect is \emph{absent} when the $b$ probe (and also
the $c$ probe) does not trigger, indicating that the conclusion in
\cite{Vdmn13}, that any particle detected in $D_1$ was earlier in $B$ (and in
$C$), is based on a misunderstanding of the important difference between a
situation in which a particular probe is triggered and when it is not
triggered. By contrast, the CH approach provides reliable quantum descriptions
both in the absence and in the presence of probes, whether or not they have
been triggered.

\xb
\outl{Vd claims regarding $w$ probe are not correct}
\xa

The claim at the end of \cite{Vdmn17}, that the weak probe $w$ employed in
\cite{Grff16} gives unreliable results, is incorrect. To begin with, the
analogy of two charged particles is misleading: a single quantum particle in a
superposition of states at two locations is not at all the same thing as two
particles, one at each location. That the $w$ probe is ``nonlocal'' does not
make it irrelevant; see Feynman's use of a long wavelength (hence ``nonlocal'')
light source in his discussion of two-slit interference in Ch.~1 of
\cite{FyLS65}. What the $w$ probe measures is the property represented by the
projector $B+C$, which in quantum mechanics, as noted above, does not mean the
same thing as ``$B$ \OR\ $C$''. Its reliability is confirmed by noting that for
particles detected in $D_3$ rather than $D_1$ the rate at which it is triggered
is exactly what one would expect. That it fails to indicate the presence of a
particle later detected by $D_1$ is just what one would expect if that particle
was never in $B+C$, as maintained in \cite{LAAZ13} and \cite{Grff16}, contrary
to \cite{Vdmn13}. What the $w$ probe does \emph{not} do, unlike $b$ or $c$, is
perturb the phase of a particle passing through the inner $B+C$ Mach-Zehnder,
and it is by ignoring this phase perturbation that \cite{Vdmn13} has arrived at
an incorrect result.

\xb
\outl{Conclusion}
\xa

In conclusion, while the phenomenological principle embodied in the notion of a
weak trace may sometimes be of value, its use in \cite{Vdmn13} and
\cite{Vdmn17} has no support in standard quantum mechanics, and can give
misleading results. By contrast, the consistent histories approach, which has
no measurement problem and has resolved numerous quantum paradoxes, provides a
consistent description of a particle passing through a nested Mach-Zehnder
interferometer, including its weak interaction with probes should some be
present.

\end{document}